\documentclass[prl,aps,floatfix,groupedaddress,showpacs,amsmath,twocolumn]{revtex4}  

\usepackage[dvips]{graphicx}
\usepackage{dcolumn}
\usepackage{bm}
\usepackage{epsfig}

\begin{document}

\title{Beyond the locality approximation in the standard diffusion Monte Carlo method}

\author{Michele Casula}

\affiliation{
Department of Physics, University of Illinois at Urbana-Champaign,
1110 W. Green St, Urbana, IL 61801, USA
} 

\date{\today}

\begin{abstract}
We present a way to include non local potentials in the standard Diffusion
Monte Carlo method without using the locality approximation. We define
a stochastic projection based on a fixed node effective
Hamiltonian, whose lowest energy is an upper bound of the true ground state
energy, even in the presence of non local operators in the Hamiltonian. The variational property
of the resulting algorithm provides a stable diffusion process,
even in the case of divergent non local potentials, like the hard-core
pseudopotentials. It turns out that the modification required to
improve the standard Diffusion Monte Carlo algorithm is simple.
\end{abstract}


\maketitle

Diffusion Monte Carlo (DMC) is one of the most successful 
methods to compute the ground state properties of quantum systems. 
Although the fixed
node (FN) approximation is needed to cure the infamous sign problem for 
fermions, the accuracy of the DMC framework 
has yielded many benchmark results\cite{foulkesreview}. However,
when the DMC method is applied to ``ab initio'' realistic Hamiltonians, 
its computational cost scales $\propto Z^{6.5}$, where $Z$ is the 
atomic number\cite{ceperley86}. 
Therefore, the use of pseudopotentials is necessary to
make those calculations feasible. 

Since the pseudopotentials are usually non
local, the ``locality approximation'' is made besides the FN, 
by replacing the true Hamiltonian $H$ with an
\emph{effective} one $H^{\mathrm{eff}}$, which reads\cite{mitas}:
\begin{equation}
H^{\mathrm{eff}} = K + V_{\mathrm{loc}} + \frac{\int dx^\prime \langle
      x^\prime | V_{\mathrm{non~loc}} | x \rangle \Psi_T(x^\prime)}{ \Psi_T(x)}, 
\label{H_locality}
\end{equation}
where $K$ is the kinetic operator, $V_{\mathrm{loc}}$ is the local potential,
and the last term in Eq.~\ref{H_locality} 
is the non local potential localized by means of the
trial wave function $\Psi_T$. The projection is then realized by iteratively 
applying the operator $G=\exp(-\tau (H^{\mathrm{eff}} - E_{\mathrm{eff}}))$ to
$\Psi_T$ in order to filter out its high energy
components. The localized potential enters in the branching part (birth and
death process) of the algorithm, 
while the usual FN constraint is employed to limit the
diffusion process within the nodal pockets of $\Psi_T$, and avoid the
fermionic sign problem. Thus $E_{\mathrm{eff}}$ is the FN ground state 
energy of $H^{\mathrm{eff}}$, computed during the sampling of the mixed
distribution $\Psi_{\mathrm{eff}} \Psi_T$:  
\begin{equation}
E_{\mathrm{eff}}=\frac{\langle \Psi_{\mathrm{eff}} | H^{\mathrm{eff}} | \Psi_T
  \rangle}{\langle \Psi_{\mathrm{eff}} | \Psi_T \rangle} = 
\frac{\langle \Psi_{\mathrm{eff}} | H | \Psi_T
  \rangle}{\langle \Psi_{\mathrm{eff}} | \Psi_T \rangle} = E_{MA}.
\label{E_MA_identity}
\end{equation}
$E_{MA}$ is the \emph{mixed average} of $H$, and the above identity holds
because $H^{\mathrm{eff}} \Psi_T/\Psi_T = H \Psi_T/\Psi_T$.
Since $\Psi_{\mathrm{eff}}$ is the FN ground state of
$H^{\mathrm{eff}}$, which differs from $H$, $E_{MA}$ 
is no longer equal to the variational FN energy of $H$, defined as:
\begin{equation}
E_{FN}=\langle \Psi_{\mathrm{eff}} | H | \Psi_{\mathrm{eff}} \rangle /
\langle \Psi_{\mathrm{eff}} | \Psi_{\mathrm{eff}} \rangle.
\label{E_FN}
\end{equation}
Therefore, in contrast with the case of local Hamiltonians, $E_{MA}$ calculated
with the locality approximation does not in general give an upper bound to the
ground state energy of $H$ (variational principle).

In a previous work\cite{lrdmc}, we introduced
the Lattice Regularized Diffusion Monte Carlo algorithm (LRDMC), which
provides an upper bound for the true ground state energy and allows estimate $E_{FN}$, 
even in the case of non local potentials. 
In this paper we propose an extension of the standard DMC framework that
gives the same results as the LRDMC method, after a proper modification of the
DMC propagator.

We start by considering the importance sampling Green function
\begin{equation}
G(x^\prime \leftarrow x, \tau) = \frac{\Psi_T(x^\prime)}{\Psi_T(x)} \langle
x^\prime | e^{-\tau (H - E_T)} | x \rangle,  
\label{green_dmc}
\end{equation}
where $E_T$ is an energy offset, $\tau$ the time step, and 
$x$ a vector of particle coordinates.
In the diffusion Monte Carlo method, $G(x^\prime
\leftarrow x, \tau)$ is iteratively applied to $\Psi_T^2$, 
in order to sample stochastically the mixed distribution $\Phi(x,t)=\Psi_T(x)
\Psi(x,t)$, $\Psi(x,t)$ converging to the lowest possible state in energy.
To rewrite $G(x^\prime \leftarrow x,\tau)$ (Eq.~\ref{green_dmc}) 
in a practical way, it is necessary
to resort to the Trotter break up, which is exact in the limit of $\tau
\rightarrow 0$. Here we split the Hamiltonian into 
local and non local operators, 
and we end up with the following expression for the Green function:
\begin{equation}
G(x^\prime \leftarrow x,\tau) \simeq \int dx'' ~ T_{x^\prime,x''}(\tau) ~
G_{DMC}(x'' \leftarrow x, \tau),
\label{new_green_function}
\end{equation}
where $G_{DMC}(x^\prime \leftarrow x, \tau)$ is the usual DMC
propagator\cite{foulkesreview},
\begin{equation}
\frac{1}{(2 \pi \tau)^{\frac{3N}{2}}} 
\exp\left[-\frac{(x^\prime - x - \tau v(x))^2}{2 \tau}\right]
e^{-\tau(E^{\mathrm{loc}}_L(x^\prime)- E_T)},
\label{DMC_green_function}
\end{equation}
and $T_{x^\prime,x}(\tau)$ is the matrix containing the non local potential,
\begin{equation}
\frac{\Psi_T(x^\prime)}{\Psi_T(x)} \langle x^\prime | e^{-\tau
V_{\mathrm{non~loc}}}| x \rangle 
  \simeq   \delta_{x^\prime,x} - \tau V_{x^\prime,x}.
\end{equation}
In the above Eqs. $N$ is the total number of particles, $v(x)=\nabla \ln
|\Psi_T(x)| $ the drift velocity,
$E^{\mathrm{loc}}_L(x)=(K+V_{\mathrm{loc}})\Psi_T(x)/\Psi_T(x)$ the
contribution to the local energy coming from the local operators, and
$V_{x^\prime,x}=\frac{\Psi_T(x^\prime)}{\Psi_T(x)} \langle
x^\prime | V_{\mathrm{non~loc}} | x \rangle$. The final form of $G_{DMC}$ has
been obtained by 
further splitting the Hamiltonian into the kinetic and potential part, while the
exponential of the non local potential in $T$ has been linearized up to
order $\tau$. 

If the case of pseudopotentials, the number of non-zero matrix elements
$V_{x^\prime,x}$ will be \emph{finite}, once a quadrature rule with a discrete
mesh of points is applied to evaluate the projection over the angular
components of the pseudopotential\cite{fahy,mitas}. Therefore, the process in 
$G(x^\prime \leftarrow x,\tau)$ driven by $T_{x^\prime,x}(\tau)$
can be calculated using a heat bath algorithm, since $T_{x^\prime,x}(\tau)
/ \sum_{x''} T_{x'',x}(\tau)$ can be seen as a transition probability, and
it can be computed \emph{a priori} for all possible new coordinates $x^\prime$.
We notice that the matrix elements $T_{x^\prime,x}(\tau)$
are easily evaluated in a standard DMC algorithm, since $V_{x^\prime,x}$
are already computed to calculate the localized pseudopotential in Eq.~\ref{H_locality}:
\begin{equation}
\frac{\int dx^\prime \langle x^\prime | V_{\mathrm{non~
      loc}} | x \rangle \Psi_T(x^\prime)}{ \Psi_T(x)} = \sum_{x^\prime} V_{x^\prime,x}. 
\label{potential_locality}
\end{equation}
At variance with the locality approximation, $ V_{x^\prime,x}$ contribute now
to move the particles, according to the transition matrix $T$ ($T$-moves).

An important limitation of this idea is given by the sign
problem. Indeed both $\frac{\Psi_T(x^\prime)}{\Psi_T(x)}$ and  $\langle
x^\prime | V_{\mathrm{non~loc}} | x \rangle$ can change sign, which should
be included in the weights, but this yields averages with
exponentially increasing noise. A solution is to apply
the FN approximation not only to $G_{DMC}$ but also to $T$, which becomes:
\begin{equation}
T^{FN}_{x^\prime,x}(\tau) = \delta_{x,x^\prime} - \tau V^-_{x^\prime,x},
\label{T_FN} 
\end{equation}
where we defined $V^\pm_{x^\prime,x}=1/2 (V_{x^\prime,x} \pm | V_{x^\prime,x} |)$. In practice,
we keep only those matrix elements which give a positive $T_{x^\prime,x}(\tau)$.
Moreover, we add to the \emph{diagonal}
potential the so called ``sign flip term'', i.e. the sum over the discarded matrix elements
$V^+_{x^\prime,x}$. Therefore, the local potential becomes
\begin{equation}
V^{\textrm{eff}}(x)=V_{\mathrm{loc}}(x) + \sum_{x^\prime} V^+_{x^\prime,x}.
\end{equation}
This is equivalent to work with a new effective FN Hamiltonian
\begin{eqnarray}
\label{H_effective}
H^{\textrm{eff}}_{x,x} & = & K + V^{\textrm{eff}}(x) \\
H^{\textrm{eff}}_{x^\prime,x} & = & \langle x^\prime | V_{\mathrm{non~loc}} | x \rangle  
\textrm{~~~if $V_{x^\prime,x}<0$}.\nonumber 
\end{eqnarray}

\begin{figure}[!htb]
\epsfxsize=85mm
\centerline{
\epsffile{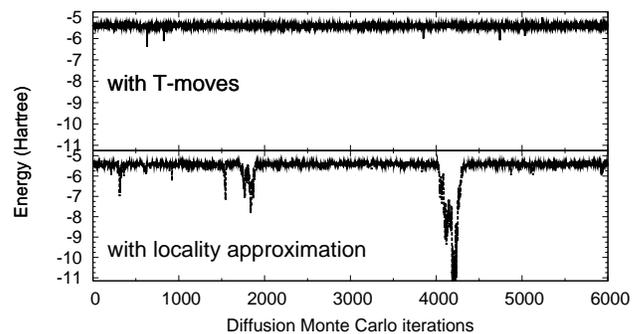}}
\caption{\label{noise}
Energies for the Carbon pseudoatom with $\tau=0.08 H^{-1}$ 
at the given DMC generation. $\Psi_T$ is an antisymmetrized geminal power (AGP)
wave function with a 3-body
Jastrow factor\cite{casula_atoms,casula_wf}. We report results for the 
locality approximation ($H^{\alpha,\gamma}$ with $\alpha=1$ and $\gamma=0$) 
and the algorithm with $T$-moves ($\alpha=0$,$\gamma=0$).
}
\end{figure}

In contrast to the effective Hamiltonian of the locality approximation 
written in Eq.~\ref{H_locality}, the
ground state energy $E_{\mathrm{eff}}(=E_{MA})$ of the above $H^{\textrm{eff}}$
is an upper bound for the ground state energy of the true $H$. 
As shown in Ref.~\onlinecite{dutch} for the Lattice Green function Monte Carlo,
this variational property is due to 
the sign flip term (\emph{positive} contribution) added to the local
potential, \emph{and} the $T$-moves driven by the off diagonal matrix elements
$V^-_{x^\prime,x}$. Instead, in the locality approximation
also $V^-_{x^\prime,x}$ is summed in the diagonal part
(Eq.~\ref{potential_locality}), and this leads to
an \emph{attractive} potential, which cannot provide a variational property for $E_{MA}$.
Moreover, we found that the negative divergences of the
fully localized potential on the nodes of $\Psi_T$ are responsible in some
case (e.g. see  
Fig.~\ref{noise}) for numerical instabilities in the locality approximation,  
which disappear once $H^{\textrm{eff}}$ in Eq.~\ref{H_effective} is used
together with the $T^{FN}$-moves. Indeed, whenever $V^-_{x^\prime,x}$ is
large, it pushes the walker away from the attractive regions of the
localized potential, and protects the sampling from divergences in the weights.  

Once a $T^{FN}$-move is generated according
to the transition probability $T^{FN}_{x^\prime,x}(\tau) / \sum_{x''}
T^{FN}_{x'',x}(\tau)$, the walker should acquire the weight $w_T(x,\tau)=\sum_{x''}
T^{FN}_{x'',x}(\tau)$ due to the normalization of the $T^{FN}$
matrix. This weight can be recast as an exponential form valid 
up to order $\tau$,
\begin{equation}
w_T=1-\tau \sum_{x^\prime} V^-_{x^\prime,x} \simeq \exp\left[-\tau \sum_{x^\prime}
  V^-_{x^\prime,x}\right].
\label{T_weight}
\end{equation} 
Thus the overall weight $w(x,\tau)$ of $G(x^\prime \leftarrow x,\tau)$ will be
\begin{equation}
w(x,\tau) = w_{DMC}~ w_T = \exp \left[-\tau (E_L(x) - E_T)
  \right], 
\label{final_weight}
\end{equation} 
where $w_{DMC}$ is the weight of $G_{DMC}$ for the
effective Hamiltonian ($V_{\mathrm{loc}}$ replaced by $V^{\textrm{eff}}$), and 
$E_L(x)=H^{\mathrm{eff}} \Psi_T/\Psi_T = H \Psi_T/\Psi_T$ is the local
energy. Notice that a non-symmetric branching factor has been included in $G_{DMC}$
(Eq.~\ref{DMC_green_function}).
When we use the exponential form in
Eq.~\ref{T_weight}, and consequently the weight in Eq.~\ref{final_weight},
the time step error is usually smaller than that obtained with the linear form.
This can be understood in the limit of perfect importance sampling.
Indeed, if $\Psi_T$ is close to the ground state of $H$, the weight 
in Eq.~\ref{final_weight} is almost constant, since the variance of $E_L(x)$ is
small, and the time step bias is reduced. 

The proposed DMC scheme for fixed node 
Hamiltonians with non local potentials is the following:
(i) perform a diffusion-drift move according to 
$G_\textrm{diff}(x^\prime \leftarrow
x, \tau)= \exp\left[-(x^\prime - x - \tau v(x))^2 /2 \tau\right]/(2 \pi
\tau)^{\frac{3N}{2}} $ \emph{as is done} in the standard
  DMC algorithm, and accept or reject this move according to the probability
\begin{equation}
min\left[1,\frac{G_\textrm{diff}(x \leftarrow x^\prime, \tau)
      \Psi_T^2(x^\prime)}{G_\textrm{diff}(x^\prime \leftarrow x, \tau)
      \Psi_T^2(x)}\right];
\label{acc_rej}
\end{equation}
(ii) weight the walker with the factor $\exp \left[-\tau (E_L(x^\prime) - E_T) \right]$;
(iii) displace the walker a second time, with a $T$-move selected
according to the transition probability $p(x'' \leftarrow x^\prime, \tau)=
T^{FN}_{x'',x^\prime}(\tau) / \sum_y T^{FN}_{y,x^\prime}(\tau)$, computed \emph{a
  priori} for all possible new $x''$.     
The branching process will be the same as in the usual DMC algorithm. In
practice, only the $T$-move is the new step, which is performed
\emph{after} weighting the walker\cite{time_step_error}. 
Although we perform an acceptance/rejection step
(Eq.~\ref{acc_rej}), which has been shown to reduce the time step
error\cite{cyrus_dmc} in $G_\textrm{DMC}$, 
the algorithm does not satisfy exactly the detailed balance
except in the limit of $\tau \rightarrow 0$, due to the break up of $G$ into
$G_\textrm{DMC}$ and $T^{FN}$ (Eq.~\ref{new_green_function}), and the use of
a non symmetric branching factor in Eq.~\ref{DMC_green_function}.

In order to estimate the variational FN energy $E_{FN}$ (Eq.~\ref{E_FN}), and
study the quality of the locality approximation, we introduce a more
general effective Hamiltonian\cite{lrdmc} $H^{\alpha,\gamma}$,
\begin{eqnarray}
\label{H_last}
H^{\alpha,\gamma}_{x,x} & = & K + V_{\mathrm{loc}}(x) + (1+\gamma) \sum_{x^\prime}
  V^+_{x^\prime,x} \nonumber \\
 & & + \alpha (1+\gamma) \sum_{x^\prime}  V^-_{x^\prime,x} \\
H^{\alpha,\gamma}_{x^\prime,x} & = & -\gamma ~ \langle x^\prime |
V_{\mathrm{non~loc}} | x \rangle  \textrm{~~~~~~~~~~~~~~~~~if $V_{x^\prime,x}>0$} \nonumber \\
H^{\alpha,\gamma}_{x^\prime,x} & = & (1-\alpha(1+\gamma))~ \langle x^\prime |
V_{\mathrm{non~loc}} | x \rangle  \textrm{~~if $V_{x^\prime,x}<0$}, \nonumber 
\end{eqnarray}
where $0 \le \alpha \le 1$ and $0 \le \gamma \le 1/\alpha -1$ are two external parameters.
In order to sample the Green function $G(x^\prime \leftarrow x, \tau)$ 
for $H^{\alpha,\gamma}$, it is sufficient to modify the matrix
$T_{x^\prime,x}(\tau)$, which becomes
\begin{equation}
T^{\alpha,\gamma}_{x^\prime,x} = \left \{
\begin{array}{ll}
1     & \textrm{if $x = x^\prime$} \label{T_matrix_new} \\
\tau ~ \gamma ~  V^+_{x^\prime,x}  & \textrm{if $V_{x^\prime,x}>0$} \\ 
- \tau ~ (1-\alpha(1+\gamma)) ~ V^-_{x^\prime,x} & \textrm{if $V_{x^\prime,x}<0$}. 
\end{array} 
\right. 
\end{equation}
The ground state $E(\alpha,\gamma)$
of $H^{\alpha,\gamma}$ is equal to $E_{MA}(\alpha,\gamma)$ (Eq.~\ref{E_MA_identity}), 
since $H^{\alpha,\gamma} \Psi_T/\Psi_T = H\Psi_T/\Psi_T$ by construction. 
The Hamiltonian in Eq.~\ref{H_effective} is recovered with $\alpha=0$ and
$\gamma=0$, while the Hamiltonian of the locality approximation
(Eq.~\ref{H_locality}) is obtained with $\alpha=1$ and $\gamma=0$. Therefore,
$H^{\alpha,\gamma}$ can interpolate between these two extremes, but the
variational principle for $E_{MA}(\alpha,\gamma)$ is not guaranteed as soon as $\alpha \ne 0$,
since the attractive term $\alpha (1+\gamma) \sum_{x^\prime}  V^-_{x^\prime,x}$
is added to the diagonal potential. However by means of $H^{\alpha,\gamma}$ one can
estimate the value of $E_{FN}(\alpha,\gamma)$ (Eq.~\ref{E_FN}), which is
\emph{variational} for every $\alpha$ and $\gamma$, since it is the expectation
value of the true $H$ on the ground state of $H^{\alpha,\gamma}$. Indeed
$H=H^{\alpha,\gamma}-(1+\gamma) \partial_\gamma H^{\alpha,\gamma}$, and the
Hellmann-Feynman theorem leads to the relation
\begin{equation}
E_{FN}(\alpha,\gamma)=E(\alpha,\gamma)-(1+\gamma)~ \partial_\gamma E(\alpha,\gamma). 
\label{evaluate_E_FN}
\end{equation}
One can show~\cite{sandro_effective} that, for a given value of $\alpha$, 
the lowest $E_{FN}(\alpha,\gamma)$ is obtained for $\gamma=0$. 
Therefore, in order to find the best variational estimate of the
ground state of $H$, it is enough 
to calculate the expression in Eq.~\ref{evaluate_E_FN} with $\gamma=0$. 
In this way one can check which $\alpha$ provides the best variational state for $H$.
The derivative $\partial_\gamma E(\alpha,0)$ can be computed with either finite
differences or correlated sampling. In both cases, one should keep in mind
that $\gamma < 1/\alpha -1$, to guarantee the positivity
of the $T^{\alpha,\gamma}$ matrix (Eq.~\ref{T_matrix_new}), 
and so calculating $E_{FN}(\alpha,0)$ becomes
harder as $\alpha$ gets closer to $1$.

Here we present the application of the method to the $Si$ and $C$ pseudoatoms. 
We computed $E_{MA}(\alpha,0)$ and
$E_{FN}(\alpha,0)$ for $\alpha=0,0.5,0.9$, and the DMC energy with the locality
approximation, which corresponds to $E_{MA}(1,0)$.
With an aim to quantify the locality error, and the correction provided by the
effective Hamiltonian $H^{\alpha,\gamma}$, we used three different trial
wave functions (with no Jastrow, a two-body, and a
three-body (electron-electron-ion) Jastrow factor respectively), 
sharing the same determinantal part, and hence the same nodes. In this way, the FN
error can be separated from the effect of the locality approximation, which causes
a dependence of the DMC energy on the \emph{shape} of $\Psi_T$.

\begin{figure}[!htb]
\epsfxsize=85mm
\centerline{
\epsfig{figure=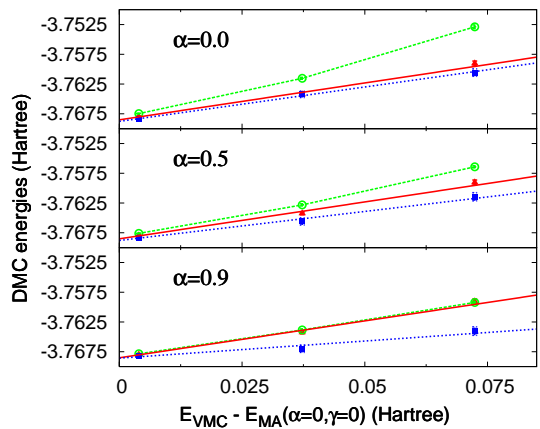,height=5.8cm}}
\caption{\label{Si_data_alpha}
(Color online) $E_{MA}(\alpha,0)$ (green, dashed line) and $E_{FN}(\alpha,0)$
  (blue, dotted line)  
  energies for the Silicon pseudoatom with different values of
  the effective Hamiltonian parameter $\alpha$. 
  The DMC energies with locality approximation
  (red, solid line), corresponding to $E_{MA}(1,0)$, are reported in all
  panels for reference. We used three different $\Psi_T$'s, which have 
  the same determinantal part. A more accurate $\Psi_T$
  corresponds to a smaller difference between its variational energy
  ($E_{VMC}$) and the $E_{MA}(0,0)$ energy, reported in abscissa.
}
\end{figure}

For the $Si$ atom we used an $s-p$ norm-conserving
Hartree-Fock (HF) pseudopotential, which is soft and  
has been generated using the Vanderbilt construction\cite{vanderbilt}.
The determinantal part of $\Psi_T$ 
is a HF wave function with a $6s6p/1s1p$ Gaussian basis set. The 2-body
Jastrow is from Ref.~\onlinecite{filippi}, while the 3-body Jastrow factor is from
Ref.~\onlinecite{casula_wf}. Both of the Jastrow factors have been optimized
using an energy minimization procedure\cite{sorella_hessian}. The results are reported in
Fig.~\ref{Si_data_alpha}. The variational $E_{FN}(\alpha,0)$ improves
going from $\alpha=0$ to $\alpha=0.9$, i.e. approaching the locality
approximation. It means that at least for this soft pseudopotential 
the locality approximation ($\alpha=1,\gamma=0$) gives
a ground state which is a good variational wave function for
$H$. 
Notice however that the standard DMC energies $E_{MA}(1,0)$
have a sizable locality error, while $E_{FN}$ with $\alpha=0.9$
depends only slightly on the shape of the trial wave function.
A similar result was obtained with
the LRDMC method for the same pseudoatom\cite{lrdmc}.

\begin{figure}[!htb]
\epsfxsize=85mm
\centerline{
\epsfig{figure=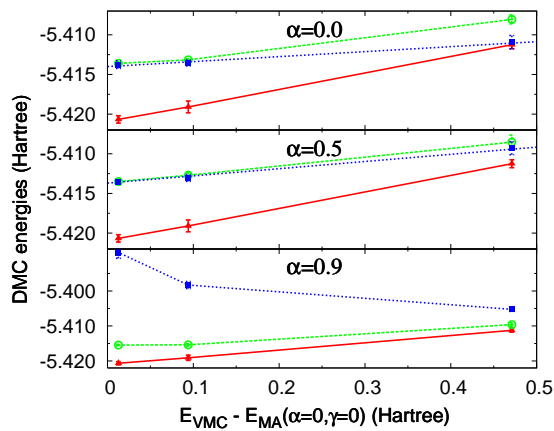,height=5.8cm}}
\caption{\label{C_data_alpha}
(Color online) The same as Fig.~\ref{Si_data_alpha}, but for the Carbon
  pseudoatom.
}
\end{figure}

For the $C$ atom we chose to work with an SBK pseudopotential\cite{SBK}, 
which is extremely hard, since it diverges like $1/r^2$ in the $s$ channel,
and $1/r$ in its local component. 
The Slater part of
$\Psi_T$ is an antisymmetrized geminal power (AGP)
wave function\cite{casula_atoms,casula_wf} 
with a $2s2p$ Gaussian basis set, optimized in the presence of the
3-body Jastrow factor by minimizing its variational energy\cite{sorella_hessian}. 
The determinantal part has been kept fixed in the other two $\Psi_T$'s, which
differ only by their Jastrow factors. The results are plotted in Fig.~\ref{C_data_alpha}.
Here the locality approximation is very poor, as it leads to 
non variational $E_{MA}$. The spikes in Fig.~\ref{noise}, 
coming from regions of the configuration space where the
\emph{effective} potential is attractive, are surely responsible of the non
variational results. Surprisingly,
$\Psi_T$ without Jastrow, which has a higher energy, leads to much
more stable DMC simulations.
The locality approximation, which relies on the quality of the 
shape of $\Psi_T$ in the core, performs poorly with this
hard-core pseudopotential, since it is difficult to find the optimal shape of
$\Psi_T$ in the core region, due to the divergence of the non local
pseudopotential. Indeed $E_{FN}(\alpha,0)$ is higher for $\alpha=0.9$, being
the worst for the 3-body Jastrow factor. On the other hand,
the best variational $E_{FN}(\alpha,0)$ is obtained for
$\alpha=0$, irrespective of the form of the Jastrow factor.

To summarize, we have described a scheme to treat non local potentials within
the standard DMC method. We have extended the DMC formalism to  
handle a generic Hamiltonian with discrete off-diagonal matrix
elements and the fixed node approximation. 
Only a simple modification of the standard algorithm is required to
include the $T$-moves generated according to the non local potentials. 
By using an effective Hamiltonian approach, we showed that it is possible to have 
stable simulations, even in the case of divergent hard-core pseudopotentials, and
obtain variational results. 
A similar effective Hamiltonian has been successfully used in the LRDMC
method. The difference is in the kinetic part, which is discretized in the
lattice regularized approach. The LRDMC and the DMC methods have
the same efficiency for small $Z$, although it is
possible to have a gain in the LRDMC efficiency by an \emph{ad hoc} choice of the
kinetic parameters, particularly for heavier elements.
We conclude, by noting that the same Green function presented here can be used in 
the Reptation Quantum Monte Carlo method\cite{baroni99}.

This work was supported by the NSF grant DMR-0404853. We thank
S. Sorella, C. Umrigar, D. M. Ceperley, S. Chiesa, and J. Kim 
for useful discussions.


\begin{thebibliography}{99}
\bibitem{time_step_error} Since the normalization of $T^\textrm{FN}_{x',x}$
  (Eq.~\ref{T_weight}) depends on $x$, the best choice is to weight the walker
  before the $T$-move.
\end{thebibliography}
\end{document}